# Supporting information for: New methods for analyzing serological data with applications to influenza surveillance

Wilfred Ndifon

Department of Immunology, The Weizmann Institute of Science, Rehovot, Israel 76100

## Summary

**For decades, the hemagglutination-inhibition (HI) assay has been used in epidemiological and basic biological studies of influenza viruses. The mechanistic basis of the assay results (called titers) is not well understood. The first part of this document describes a biophysical model of HI that illuminates the mechanistic basis of and provides the theoretical motivation for new ways of interpreting titers. The biophysical model is applicable to other serological assays; this fact is illustrated using the neutralization assay. The second part of the document describes some new ways of interpreting titers, which involve, among other methods, singular value decomposition and probabilistic multidimensional scaling. The third part of the document discusses biological and mathematical issues related to the determination of the effective dimensionality of titers, and describes an algorithm for recovering unavailable titers.**

## Part 1. Biophysical model of HI

**1.1** *Motivation for the model*

Presently, there is a paucity of reliable quantitative information about the biophysical determinants of titers. More specifically, there is inadequate knowledge about the manner in which both antigenic and non-antigenic parameters contribute to titers. Such knowledge would not only allow comparisons of titers obtained under different experimental conditions (e.g., using different concentrations of virus and of red cell), it would also enable critical examination of existing approaches to interpreting titers. While there are computational methods for inferring antigenic relationships from titers [1], to our knowledge there has been only one serious attempt [2] at shedding light on the nature of the biophysical interactions that determine these titers. In particular, Lanni and Lanni [2] investigated the kinetics of HI and derived mathematical equations for the dynamics of the concentration of, for example, virus-antibody and virus-red cell complexes. However, those authors did not derive an equation for the titer. Here, such an equation is derived. The approach presented here differs from that of Lanni and Lanni in important ways. Firstly, an equilibrium (as opposed to a non equilibrium) chemical kinetic framework is used, which is motivated by previous experimental results [3-7] suggesting that the virus-antibody and virus-red cell interactions that underlie HI are similar in nature to more general enzyme-substrate interactions, and that those HI interactions can attain the equilibrium state[1] on time scales that are typical of the HI assay. Secondly, a collision theory [8] approach is employed, allowing virus-antibody and virus-red cell complexes of arbitrary sizes to be accounted for, in contrast to the absolute rate theory approach employed by Lanni and Lanni. Thirdly, the emphasis is on shedding light on basic aspects of HI; therefore, a smaller number of simplifying assumptions are made concerning the nature of HI.

More specifically, the analyses described here pertain to an HI assay that is designed to quantify the titer of virus X relative to antibody-containing serum (or "antiserum") raised against virus Y, denoted $H^{XY}$. The analyses make possible the derivation of a mathematical equation for $H^{XY}$. This mathematical equation is used subsequently to investigate the accuracy of commonly-used quantitative measures of the antigenic similarity[2] of virus *X* to virus *Y*, namely the normalized titer of virus *X* relative to virus *Y*-derived antiserum and the Archetti-Horsfall measure (denoted AHM) of the antigenic similarity between viruses *X* and *Y*.

---

[1]Although the assumption that HI interactions attain equilibrium is consistent with previous experimental results (e.g., see [6]), it should be noted that the attainment of equilibrium is dependent on the HI assay conditions (e.g., pH, temperature). Moreover, bound influenza viruses can elute from a red cell by cleaving receptors found on the red cell surface [9]. This raises the possibility that as influenza viruses continuously bind to and elute from cell surface receptors, those receptors could eventually become depleted, thereby rendering red cells non-agglutinable [4,10]. However, given that there is on the order of a thousand receptors on each red cell surface [11,12] such loss of agglutinability may occur on time scales (respectively temperatures) that are much longer (respectively higher) than those typical of HI assays [10].

[2]The antigenic similarity of virus *X* to virus *Y* is defined as the degree to which antibodies raised against virus *Y* neutralize virus *X*. Note that antigenic similarity, as defined here, is not necessarily symmetric.

## 1.2. *Preliminaries*

The main steps involved in determining the titer of virus *X* relative to virus *Y*-derived antiserum are as follows (e.g., see [13]): Firstly, virus *X* is propagated in chicken eggs (or in cell culture), yielding a concentrated solution of the virus. Serial 2-fold dilutions of the virus solution are made and a fixed volume of each diluted solution is mixed with a fixed amount of red cells (see Figure S1). The virus-red cell mixtures are incubated for a specified amount of time (e.g., 1 hour) and then examined for the occurrence of agglutination, which is characterized by the absence of a "button" at the bottom of the experimental vessel. The highest dilution of the virus solution that elicits marked agglutination of red cells is called the hemagglutination endpoint. The corresponding amount of virus is denoted by 1 hemagglutinating unit. After determining the hemagglutinating unit of virus *X*, 10 serial 2-fold dilutions of antiserum obtained from organisms (e.g., ferrets) infected by virus *Y* are made. A fixed volume of each diluted antiserum is mixed with an equal volume of a solution containing 4 hemagglutinating units of virus *X*. To this antibody-virus mixture is then added a fixed amount of red cells (the same as the amount of red cells used in determining the hemagglutinating unit, as mentioned above). The 10 antibody-virus-red cell mixtures thus obtained are incubated, and checked subsequently for the occurrence of agglutination. The titer is defined as the reciprocal of the highest dilution of antiserum that inhibits marked agglutination of the red cells.

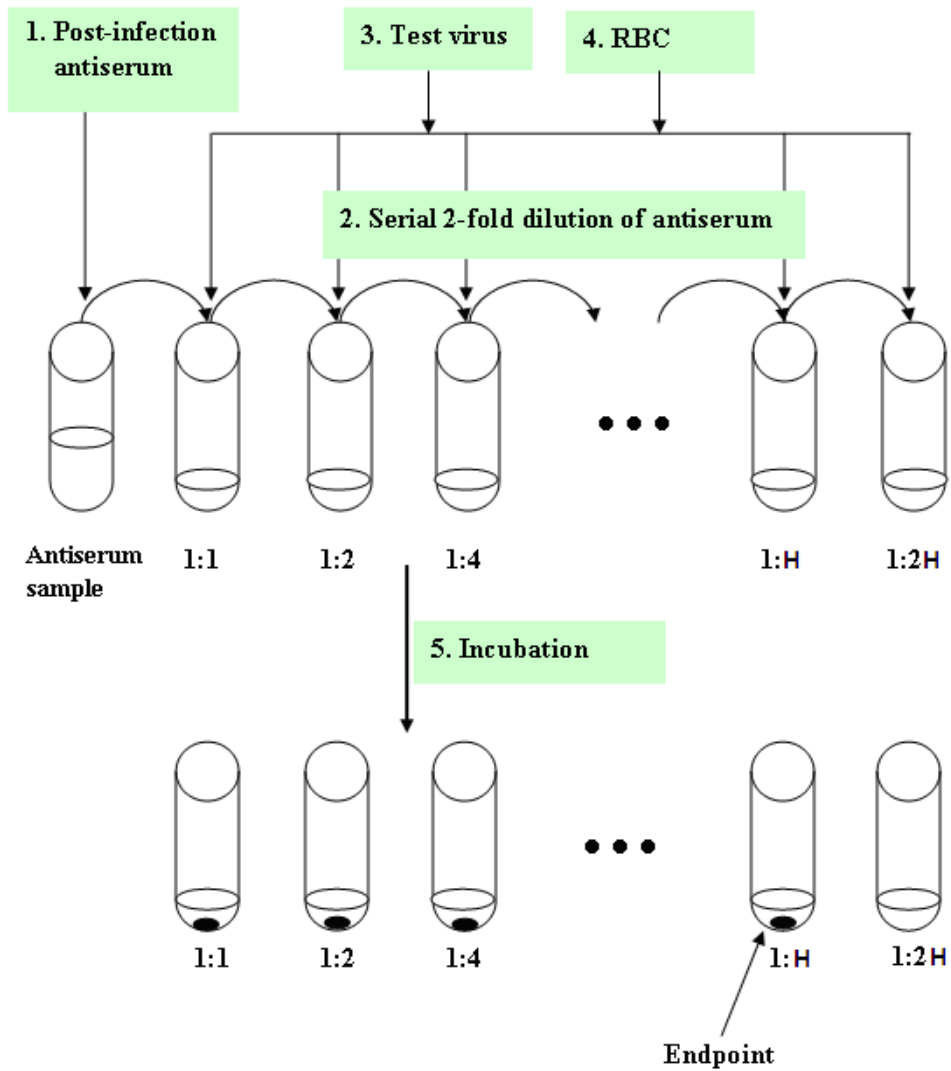

**Figure 1. Schematic illustration of the main steps of the HI assay.** Antisera taken from experimentally infected ferrets, or from other sources, are diluted serially in microtiter plates (test tubes are shown for ease of presentation). The virus under investigation is grown in cell culture or in hen's eggs to produce a concentrated virus solution. A standardized amount of the virus (see text) is then added to the diluted antisera followed by addition of a suspension of red cells. The resulting mixtures are incubated for a fixed amount of time (e.g., 1 hour). Following incubation, the mixtures are examined for the presence of hemagglutination, characterized by the absence of a button at the bottom of the plate. The HI titer is defined as the reciprocal of the highest dilution of antiserum that inhibits agglutination. In the example shown, the HI titer is given by $A_0/A_s=H$, where $A_s$ is the concentration of antibodies found in the 1:H dilution and $A_0$ is the concentration of antibodies found in undiluted antiserum.

Note that antisera generally contain different populations of antibodies each of which may have a different affinity for the target virus. The measured affinity of antisera for a virus is the average of the affinities of these different antibody populations. In addition, note that the HI assay protocol is similar to protocols for other serological assays, including the more sensitive neutralization assay [13]. The main difference between the protocols for the HI and neutralization assays is that in the latter assay samples taken from equilibrium antibody-virus mixtures are incubated with susceptible target cells (e.g., Mardin-Darby canine kidney cells), and the neutralization titer is defined as the reciprocal of the maximum dilution of antiserum that inhibits infection in at least 50% of the antibody-virus-cell mixtures.

In the following, the various steps of the HI assay described above are explicitly modeled. First, the kinetics of agglutination of red cells by influenza viruses are analyzed, followed by the kinetics of antibody-mediated inhibition of agglutination. Connections to the neutralization assay are subsequently discussed.

### 1.3. *Agglutination of red cells by influenza viruses*

The surface of each influenza virus contains numerous hemagglutinin (HA) molecules [9,14]. Each HA molecule contains a specific site (also called a receptor binding site) that can interact with and bind to a cognate sialic acid-containing receptor found on the surface of a red cell. Typically, multiple receptor binding sites found on the surface of a given virus bind to distinct cell surface receptors found on the same red cell surface; each virus-red cell "bond" is multivalent[3]. This multivalency is necessary in order to ensure the stability of a virus-red cell bond since individual receptor binding site-cell surface receptor bonds are very weak, having equilibrium dissociation constants that are on the order of a millimolar [15]. Let *a* denote the maximum number of red cells that can be bound simultaneously by the same influenza virus (i.e., the "effective" number of receptor binding sites)[4] and let $b$ denote the maximum number of influenza viruses that can bind simultaneously to the same red cell (i.e., the "effective" number of cell surface receptors). Previous experimental results suggest that $a \approx 2$ [16]. Meanwhile, for chicken red cells, experimental estimates of $b$ range from 830 to 1660 [12]. Higher estimates of $b$ (up to 5,000) have been obtained, using influenza B viruses [11].

---

[3] The formation of the first receptor binding site-cell surface receptor bond between a virus and a red cell brings adjacent receptor binding sites (found on the surface of the virus) and cell surface receptors (found on the surface of the red cell) closer to each other and, hence, facilitates the formation of additional receptor binding site-cell surface receptor bonds. Accordingly, the rate of formation of a multivalent virus-red cell bond would be given approximately by the rate of formation of the first receptor binding site-cell surface receptor bond.

As noted earlier, the titer is determined experimentally by observing the pattern of agglutination occurring in 10 mixtures, each containing fixed amounts of virus and of red cell and differing amounts of antibody-containing antiserum. Consider the interactions between virus and red cell occurring in the $i$th mixture, $1 \leq i \leq 10$. Let $V$ denote the molar concentration of virus particles found in this mixture, $B$ the concentration of red cells, $A_i$ the concentration of antibodies, and $x_i$ the concentration of virus-red cell bonds found in the $i$th virus-antibody-red cell mixture. The collision frequency $Z_{BV}$ between virus and red cell can be estimated from first principles [8]. During a particular collision between virus and red cell the probability that a receptor binding site of the virus is free is given by $(aV - x_i)/aV$, while the probability that a cell surface receptor of the red cell is unoccupied is given by $(bB - x_i)/bB$. Now, let $\rho$ denote the probability that during the collision a free receptor binding site will make contact with an unoccupied cell surface receptor, in the proper orientation (i.e., such that steric hindrance is overcome). Based on the Maxwell-Boltzmann distribution of the kinetic energies of the colliding particles a fraction, $\exp(-E_a/RT)$, of such properly oriented collisions will result in the formation of a virus-red cell bond, where $E_a$ is the activation energy for virus-red cell bond formation. On the other hand, virus-red cell bonds will dissociate at the rate $f \exp(-E_d/RT)$, where $E_d$ is the activation energy for dissociation of a virus-red cell bond and $f$ is a pre-exponential factor (this dissociation rate depends on the activity of the influenza viral neuraminidase protein). Therefore, ignoring, for now, the presence of antibodies, the dynamics of the concentration of virus-red cell bonds is given by[5]:

$$\frac{dx_i}{dt} = \left(\frac{2V - x_i}{2V}\right)\left(\frac{bB - x_i}{bB}\right) BVZ_{BV}\rho \exp(-E_f/RT) - f\exp(-E_d/RT)x_i$$

$$= K_1 \frac{(2V - x_i)(bB - x_i)}{2b} - K_2 x_i, \qquad (S1)$$

where $K_1 = Z_{BV}\rho(-E_a/RT)$ and $K_2 = f\exp(-E_d/RT)$. Observe that $a=2$ in (S1).

---

[4] The significantly larger size of a red cell compared to that of an influenza virus - the volume of a spherical influenza virus particle, with a diameter of $10^{-1} \mu m$ [15], is $10^{-4} \mu m^3$, while that of a chicken red cell is estimated to be $10^2 \mu m^3$ [16] - would limit severely the number of red cells that can be bound simultaneously by the same influenza virus.

Let $x_i^e$ denote the equilibrium concentration of virus-red cell bonds found in the $i$th virus-antibody-red cell mixture. At equilibrium, the concentration of doubly-bound viruses is $(x_i^e)^2/4V$ and the fraction of agglutinated red cells (i.e., the fraction of red cells bound by a doubly-bound virus) is given by:

$$f_{x_i^e} = \frac{(x_i^e)^2}{2BV}. \qquad (S2)$$

### 1.4. *Antibody-mediated inhibition of agglutination*

Antibodies recognize and bind to specific antigenic sites of HA molecules found on the influenza viral surface. Theoretical interpretations [5,17] of experimental data on influenza virus-antibody interactions suggest that there are ~2000 antigenic sites per influenza virus. The mechanism by which antibodies neutralize[6] influenza viruses and, hence, inhibit red cell-agglutination [18] is not well understood. The physical closeness of antigenic sites to receptor binding sites and the relatively large sizes of antibodies suggest that bound antibodies may neutralize influenza viruses by occluding receptor binding sites, thereby preventing those sites from binding to cell surface receptors [19,20]. This "occlusion" mechanism is supported by the results of previous experimental studies [7,21,22] on antibody-mediated neutralization of influenza viruses. Those experimental results suggest that the fraction of neutralized influenza viruses increases with the average number of bound antibodies per virus in a sigmoidal manner. The occlusion mechanism described above could be relevant for the neutralization of a number of other viruses, including the human immunodeficiency virus (reviewed in [20]).

---

[5] A virus can only form one virus-red cell bond with a particular red cell. Since the number of red cells is typically very large (see arguments presented later in the text) it is not necessary to account explicitly for the (negligible) probability that the virus and red cell under consideration are not already bound to each other.

[6] When we talk of neutralization, we are referring specifically to neutralization of infectivity, which is a sufficient, although not a necessary, condition for inhibition of hemagglutination [19].

Note that in spite of the high average number of antibodies that were bound to each neutralized influenza virus at the neutralization endpoint [7,20,21], the kinetics of neutralization was apparently first-order [7]. This led to the hypothesis that neutralization may require that a single antibody binds to an antigenic site that is "critical" or "neutralization-relevant" [7]. According to this hypothesized "critical sites" mechanism, when an antibody binds to a critical antigenic site it may elicit conformational (and other types of) changes to the entire viral surface, and those changes may render receptor binding sites incapable of binding to cell surface receptors [20]. However, there is no documented evidence of such changes to the influenza viral surface induced by antibody binding. Moreover, as has been noted on many occasions (e.g., [23]), the observed pseudo-first-order kinetics of neutralization could be due to the fact that antibodies were used in excess over influenza viruses in the cited experiments. In light of the paucity of supporting experimental evidence, the critical sites mechanism will not be considered further.

The occlusion mechanism of antibody-mediated neutralization of influenza viruses suggests that the probability of neutralization could be described mathematically by the following equation:

$$p(\bar{y}) = \frac{\bar{y}^h}{\lambda^h + \bar{y}^h}, \tag{S3}$$

where $\bar{y}$ is the average number of bound antibodies per virus, $\lambda$ is the average number of bound antibodies per virus at 50% neutralization, and $h$ is the Hill coefficient. For the HC10 antibody used in [7] the Hill coefficient is $h \approx 3$ [24].

If $y_i$ denotes the concentration of virus-antibody bonds found in the $i$th antibody-virus-red cell mixture, then the average number of bound antibodies per virus is given by $y_i/V$. Therefore, substituting $y_i/V$ for $\bar{y}$ in (S3) gives the probability that a particular virus is neutralized:

$$p\left(\frac{y_i}{V}\right) = \frac{y_i^h}{(\lambda V)^h + y_i^h}. \tag{S4}$$

Modifying eqn. (S1) to account for the fact that only un-neutralized viruses can bind red cells gives the following equation for the dynamics of the concentration of virus-red cell bonds:

$$\frac{dx_i}{dt} = K_1 \frac{(2V - x_i)(bB - x_i)}{2b} \frac{(\lambda V)^h}{(\lambda V)^h + y_i^h} - K_2 x_i. \tag{S5}$$

At equilibrium, the concentration of virus-red cell bonds found in the *i*th virus-antibody-red cell mixture is given by:

$$x_i^e = \frac{\varphi \pm \left(\varphi^2 - 8bBK_a^2 V(\lambda V)^{2h}\right)^{1/2}}{2K_a(\lambda V)^h}, \tag{S6}$$

where $\varphi = (\lambda V)^h (2b + bBK_a + 2K_a V) + 2b(y_i^e)^h$ and $K_a = K_1/K_2$ is the equilibrium association constant for virus-red cell binding, that is, the affinity of virus for red cell.

A corresponding equation for the dynamics of the concentration of virus-antibody bonds can be derived using the same principles as those outlined above. Briefly, viruses will collide with antibodies with frequency $Z_{AV}$ [8]. Let $n$ denotes the number of antigenic sites per virus. The probability that an antigenic site is unoccupied is given by $(nV - y_i)/nV$. Meanwhile, the probability that an antibody is free is given by $(A_i - y_i)/A_i$. As before, let $\tilde{\rho}$ denote the probability that during a given collision between virus and antibody the antibody will make contact with a free antigenic site of the virus in the proper orientation. A fraction, $\exp(-\tilde{E}_a/RT)$, of such properly oriented collisions will result in the formation of a virus-antibody bond, where $\tilde{E}_a$ is the activation energy for virus-antibody bond formation. On the other hand, virus-antibody bonds will dissociate at the rate $\tilde{f} \exp(-\tilde{E}_d/RT)$, where $\tilde{E}_d$ is the activation energy for dissociation of a virus-antibody bond and $\tilde{f}$ is a pre-exponential factor. Therefore, the dynamics of the concentration of virus-antibody bonds is given by:

$$\frac{dy_i}{dt} = \tilde{K}_1 \frac{(nV - y_i)}{n}(A_i - y_i) - \tilde{K}_2 y_i, \tag{S7}$$

where $\tilde{K}_1 = Z_{AV}\tilde{\rho}\exp(-\tilde{E}_a/RT)$ and $\tilde{K}_2 = \tilde{f}\exp(-\tilde{E}_d/RT)$.

Due to the large size of a red cell, each virus-red cell bond would occlude a certain number of antigenic sites on the bound virus. Let us assume that, as in the case of viral neutralization by antibodies, the probability that a given antigenic site is occluded by red cells is an increasing, sigmoidal function of the average occupancy of receptor binding sites, with a maximal probability of $\alpha$, $0 \leq \alpha \leq 1$, and a half-maximal average occupancy of 1:

$$p\left(\frac{x_i}{V}\right) = \frac{\alpha x_i}{V + x_i}. \tag{S8}$$

Using (S7) and (S8), the equation for the dynamics of the concentration of virus-antibody bonds becomes:

$$\frac{dy_i}{dt} = \tilde{K}_1 \frac{(nV - y_i)}{n}(A_i - y_i)\left(\frac{V + (1-\alpha)x_i}{V + x_i}\right) - \tilde{K}_2 y_i. \tag{S9}$$

At equilibrium, the concentration of virus-antibody bonds found in the $i$th virus-antibody-red cell mixture is given by:

$$y_i^e = \frac{\phi \pm \left(\phi^2 - 4\tilde{K}_a^2 nA_0 V(V + (1-\alpha)x_i^e)^2\right)^{1/2}}{2\tilde{K}_a(V + (1-\alpha)x_i^e)}, \tag{S10}$$

where $\phi = \tilde{K}_a A_0(V + (1-\alpha)x_i^e) + nV(1 + \tilde{K}_a(V + (1-\alpha)x_i^e)) + nx_i^e$ and $\tilde{K}_a = \tilde{K}_1/\tilde{K}_2$ is the equilibrium association constant for virus-antibody binding, that is, the affinity of antibody for virus.

### 1.5. *Mathematical definition of the titer*

As noted earlier, the titer of virus *X* relative to antiserum containing antibodies raised against virus *Y* (i.e., $H^{XY}$) is defined as the reciprocal of the highest dilution of the antiserum that inhibits marked agglutination of red cells by virus *X*; that is, the ratio of the concentration of antibodies found in undiluted antiserum to the minimum concentration of antibodies that inhibits marked agglutination (see Figure S1). If we let *l*, $0 \leq l \leq 1$, denote a threshold value indicating marked agglutination (e.g., 70% agglutination), then $H^{XY}$ is defined mathematically as follows:

$$H^{XY} = \frac{A_0}{\min\{A_i : f_{x_i^e} \leq l\}} = \frac{A_0}{\min\{A_i : x_i^e \leq \sqrt{2lBV}\}}, \tag{S11}$$

where $A_0$ is the concentration of antibodies found in undiluted antiserum and $f_{x_i^e}$ is given by (S2).

A straightforward approach to obtaining a mathematical expression for $H^{XY}$ is to solve (S5) and (S10) for $x_i^e$; set $x_i^e \leq \sqrt{2lBV}$ and solve for $A_i$; and set $H^{XY} = A_0/\min\{A_i\}$. However, this approach is analytically intractable. Therefore, the following approach is used to obtain an approximate expression for $H^{XY}$: substituting for $y_i^e$ in $dx_i/dt = 0$ and solving for $A_i$ gives:

$$A_i = \frac{\lambda \tilde{\gamma}^{1/h} V}{\tilde{K}_a}\left(\tilde{K}_a + \frac{n(V + x_i^e)}{V(V + (1-\alpha)x_i^e)(2^{1/h}n - \lambda\tilde{\gamma}^{1/h})}\right), \tag{S12}$$

where

$$\tilde{\gamma} = \frac{K_a(2V - x_i^e)(bB - x_i^e) - 2bx_i^e}{2bx_i^e}. \tag{S13}$$

Now, substituting $x_i^e \leq \sqrt{2lBV}$ into (S12) gives:

$$A_i \geq \frac{\lambda \gamma^{1/h} V}{\tilde{K}_a}\left(\tilde{K}_a + \frac{n(V + \sqrt{2lBV})}{V(V + (1-\alpha)\sqrt{2lBV})(2^{1/h}n - \lambda\gamma^{1/h})}\right), \tag{S14}$$

where

$$\gamma = \frac{K_a(2V - \sqrt{2lBV})(bB - \sqrt{2lBV}) - 2b\sqrt{2lBV}}{2b\sqrt{2lBV}}. \tag{S15}$$

It follows from (S11) and (S14) that the value of $A_i$ that satisfies $\min\{A_i : x_i^e \leq \sqrt{2lBV}\}$ is given approximately by: $\lambda\gamma^{1/h}V(\tilde{K}_a + n(V + \sqrt{2lBV})/V(V + (1-\alpha)\sqrt{2lBV})(2^{1/h}n - \lambda\gamma^{1/h}))/\tilde{K}_a$, and therefore

$$H^{XY} \approx \frac{\tilde{K}_a A_0}{\lambda\gamma^{1/h}V(\tilde{K}_a + n(V + \sqrt{2lBV})/V(V + (1-\alpha)\sqrt{2lBV})(2^{1/h}n - \lambda\gamma^{1/h}))}. \tag{S16}$$

Note that (S16) is an approximation because the expression for $\min\{A_i : x_i^e \leq \sqrt{2lBV}\}$ is also an approximation. It is important to bear in mind that the accuracy of the experimentally measured value of $H^{XY}$ depends crucially on the dilution factor used in the HI assay. For greater accuracy the dilution factor should be very close to 1. However, this is impractical for many reasons, including the fact that an astronomical number of antiserum dilutions may be required in order for the HI assay endpoint to be reached.

It is worth drawing attention to an interesting connection between the general mathematical form of the HI equation derived above and the mathematical form of the equation for the neutralization titer. As discussed earlier, the neutralization titer is defined as the reciprocal of the maximum dilution of antiserum raised against virus $Y$ (or the minimum concentration of antibodies) that prevents infection of susceptible cells by virus $X$ in at least 50% of cell cultures inoculated with a sample taken from a mixture of the antiserum and virus $X$ [11]. Let $c$ be the minimum concentration of free or un-neutralized particles of virus $X$ required to infect at least one cell in 50% of inoculated cell cultures. The dissociation of bonds between antibodies and virus X following inoculation into cell cultures would occur very infrequently since the concentration of virus particles used (and, hence, the maximum possible concentration of antibody-virus bonds) is very low [11], while $\tilde{K}_2$, the dissociation rate of an antibody-virus bond, is typically small. Therefore, viruses that are neutralized prior to inoculation could remain in that state, while un-neutralized viruses would mostly account for subsequent infection of susceptible cells. The neutralization titer can be defined as:

$$N^{XY} \approx \frac{A_0}{\min\{A_i : p \geq 1 - c/(rV)\}}, \tag{S17}$$

where $p$, the equilibrium fraction of neutralized viruses, is given by (4), and $r$ is the ratio of infectious to non-infectious viral particles (for influenza viruses, $r \sim .02$ [14]).

It can be seen from both (S4) and (S7) that $p \geq 1 - c/(rV)$ implies:

$$\tilde{K}_a(nV - \varphi)(A_i - \varphi) \geq \varphi, \tag{S18}$$

where

$$\varphi = \lambda V \left(\frac{rV}{c} - 1\right)^{1/h}. \tag{S19}$$

Using (17) and (18) we find that the neutralization titer is given by the expression:

$$N^{XY} \approx \frac{\tilde{K}_a A_0}{\varphi\left(\tilde{K}_a + n/(nV - \varphi)\right)}, \tag{S20}$$

which has a similar mathematical form as (S16).

### 1.6. *Some remarks on the derived HI equation*

Recall that in (S16), which quantifies the titer of virus *Y* relative to virus *X*-derived antiserum, the parameter *V* denotes the concentration of virus *X*, *R* the concentration of red cells (which is typically standardized across HI assays), $K_a$ the avidity of virus *X* for red cell, $A_0$ the concentration of antibodies found in virus *Y*-derived antiserum, and $\tilde{K}_a$ the affinity of those antibodies for virus *X*. For clarity, super-scripts will be used to further specify these parameters; for example, $A_0^Y$ will be used to denote the concentration of antibodies found in virus *Y*-derived antiserum and $\tilde{K}_a^{XY}$ to denote the affinity of those antibodies for virus *X*. In the following, Eqn. (S16) is used to assess the accuracy of two commonly-used HI-based methods of quantifying the antigenic difference between influenza viruses. First, some observations on the mathematical form of (S16) are in order. Eqn. (S16) makes a number of parameter-free predictions that are intuitive and also consistent with empirical data. Firstly, as expected (e.g., see Tables I & II in [25]), the equation predicts that the titer of virus *X* relative to virus *Y*-derived antiserum, $H^{XY}$, increases linearly with the concentration of antibodies found in the antiserum, $A_0^Y$, and it also increases with the affinity of those antibodies for virus *Y*, $\tilde{K}_a^{XY}$, and with the agglutination threshold, *l*.

On the other hand, $H^{XY}$ is predicted to decrease with the concentration of virus *X*, $V^X$, with the avidity of virus *X* for red cell, $K_a^X$, and also with the average number of antibodies that must be bound to each particle of virus *X* in order to neutralize it with a probability of 50% (i.e., $\lambda$). Indeed, the predicted inverse correlation between $H^{XY}$ and $K_a^X$ is consistent with experimental data showing that the titer decreases as the strength of virus-red cell interactions increases, due to passage of virus in eggs [4]. It is also worth noting that eqn. (S16) predicts that $H^{XY}$ would be approximately invariant to changes in $\tilde{K}_a^{XY}$ if

$$r = n\left(V^X + \sqrt{2lBV}\right) / \left(V^X\left(V^X + (1-\alpha)\sqrt{2lBV}\right)\left(2^{1/h}n - \lambda\gamma^{1/h}\right)\right) << \tilde{K}_a^{XY}. \tag{S21}$$

Meanwhile, $H^{XY}$ would increase approximately linearly with $\tilde{K}_a^{XY}$ if $r >> \tilde{K}_a^{XY}$, and it would increase with $\tilde{K}_a^{XY}$ in a sigmoidal manner for intermediate values of $r$. Therefore, the sensitivity of the HI assay to antigenic differences between influenza viruses could be influenced by the value of $r$. Below, possible ranges for $r$ are estimated using reasonable values of relevant experimental parameters, providing insight into the potential sensitivity of the HI assay.

Consider the virus-red cell mixture that corresponds to the hemagglutination endpoint (see above) and denote by $\tilde{V}^X$ the concentration of virus particles (i.e., 1 hemagglutinating unit) found in this mixture and by $\tilde{R}$ the concentration of red cells. The virus-red cell mixture contains 0.5ml of a .5% standardized suspension of red cells in a 1ml solution [26]. In the case of chicken red cells, 0.5ml of a .5% standardized chicken red cell suspension contains $3\times10^7$ red cells [27]. In addition, for influenza viruses that are pre-adapted to red cells (i.e., by means of egg passage), the virus-red cell mixture typically [28,29] contains approximately as many viruses as red cells[7]. In other words, the concentration of virus particles found in the virus-red cell mixture is given approximately by $\tilde{V}^X \approx (3\times10^7 \times 10^3)/(6.02\times10^{23}) = 5\times10^{-14} M$, suggesting that the concentration of virus particles used in the HI assay is $V^X = 4\times\tilde{V}^X \approx 2\times10^{-13} M$.

On the other hand, $\tilde{K}_a^{XY}$ was found to range from $6\times10^6$ to $1\times10^9 M^{-1}$ [30], for monoclonal antibodies targeted to all five HA epitopes of influenza viruses (and their antibody escape mutants) belonging to the H1N1 and H3N2 subtypes; and from $7\times10^8$ to $2\times10^9 M^{-1}$ [31], for monoclonal antibodies targeted to three HA epitopes of an influenza virus belonging to the H7N1 subtype[8]. Using $\tilde{K}_a^{XY} = 1 \times 10^9 M^{-1}$, $V^X \leq 1 \times 10^{-13} M$, $n=2000$ [18], and $h=3$ [24], it can readily be shown that $r >> \tilde{K}_a^{XY}$ for a wide range of values of both $\alpha$ and $\lambda\gamma^{1/h}$. Therefore, the HI assay could be fairly sensitive to antigenic differences between influenza viruses. Indeed, the HI assay can detect immunologically consequential point mutations to individual influenza virus HA epitopes much more accurately than sophisticated methods such as the enzyme-linked immunosorbent assay [30].

---

[7] At the hemagglutination endpoint the ratio of the number of red cell-associated influenza viruses to the number of red cells is estimated to be ~1 [29,32]. Since virus-red cell bonds are extremely stable [30], the vast majority of viruses found at the hemagglutination endpoint would be associated with red cells, hence the conclusion that 1 hemagglutinating unit contains approximately the same number of viruses as it does red cells.

The foregoing discussion suggests that it is reasonable to approximate $H^{XY}$ by:

$$H^{XY} \approx \frac{\tilde{K}_a^{XY} A_0^Y V^X \left(V^X + (1-\alpha)\sqrt{2lBV^X}\right)\left(2^{1/h}n - \lambda\gamma^{1/h}\right)}{\lambda\gamma^{1/h} n \left(V^X + \sqrt{2lBV^X}\right)}. \quad (S22)$$

Eqn. (S22) predicts that $H^{XY}$ can be written as the product of $C^{XY} = A_0^Y \tilde{K}_a^{XY}$, which depends on the antigenic difference between viruses $X$ and $Y$, and $J^X = V^X \left(V^X + (1-\alpha)\sqrt{2lBV^X}\right)\left(\sqrt{2}n - \lambda\gamma^{1/h}\right) / \left(\lambda\gamma^{1/h} n \left(V^X + \sqrt{2lBV^X}\right)\right)$, which depends on such non-antigenic parameters as the amount of virus $X$ and the avidity of virus $X$ for red cell:

$$H^{XY} \approx A_0^Y \tilde{K}_a^{XY} J^X \quad (S23)$$

This suggests that a natural way to decouple the antigenic and non-antigenic contributions to $H^{XY}$ is to transform the titer logarithmically, since the log-transformed titer is additive in the effects of antigenic and non-antigenic variables.

### 1.7. *Accuracy of existing measures of antigenic difference*

The normalized titer (NHT) of virus $X$ relative to virus $Y$-derived antiserum is the most commonly-used measure of the antigenic similarity of virus $X$ to virus $Y$ (e.g., [13,33]). More specifically, NHT is defined as:

$$NHT^{XY} = \frac{H^{XY}}{H^{YY}} = \frac{C^{XY} J^X}{C^{YY} J^Y}. \quad (S24)$$

---

[8] The (polyclonal) antibodies found in antisera are heterogeneous with respect to their avidity for virus and the average affinity of those antibodies for a given virus strain could be lower than the affinity of a monoclonal antibody targeted to a dominant HA epitope of that virus strain.

Observe that for a given amount of antigenic difference between X and Y, $H^{XY}$ would increase with the amount of antibodies found in antisera, which varies with the ability of virus $Y$ to induce antibodies in infected hosts as well as on the immune status of those hosts [34]. Normalizing $H^{XY}$ by $H^{YY}$ is useful because it helps to control for the potential confounding effect of the amount of antibodies on the estimation of antigenic differences. Unfortunately, as indicated by the presence of both $J^X$ and $J^Y$ in (S24), the normalized titer still depends on non-antigenic variables, which may confound estimates of the antigenic difference between viruses $X$ and $Y$. For example, if $J^X$ is much smaller than $J^Y$ (as may occur if virus $Y$ has much lower avidity for red cell than virus $X$), then the normalized titer may predict, incorrectly, that virus $X$ is antigenically very different from virus $Y$ even if both viruses are antigenically similar.

In contrast to the normalized titer, the AHM of the antigenic difference between viruses $X$ and $Y$ is defined only for cases when both homologous and heterologous titers are available for the two viruses. More precisely, the AHM is given by [36,37]:

$$AHM^{XY} = AHM^{YX} = \left(\frac{H^{XY}}{H^{YY}}\frac{H^{YX}}{H^{XX}}\right)^{1/2} \approx \left(\frac{C^{XY}C^{YX}}{C^{XX}C^{YY}}\right)^{1/2}. \tag{S25}$$

Eqn. (S25) predicts that the AHM is approximately independent of the non-antigenic variables represented by both $J^X$ and $J^Y$. In other words, the AHM is influenced by non antigenic variables to a lesser degree than is the normalized titer; the AHM could afford a more reliable measure of antigenic differences between viruses than the normalized titer. This is consistent with results presented in [38]. The above results suggest that AHM should be preferred over the normalized titer, whenever possible. This could be particularly beneficial when there is significant variation in the non-antigenic properties of the viruses being analyzed.

## Part 2. Some applications of the biophysical model

### 2.1. *Singular value decomposition (SVD) of titers*

In mathematical terms, an HI table $H$ has the form of an $m$ by $n$ matrix, with $m$ the number of viruses represented in the table and $n$ the number of antisera. Each entry $H^{XY}$ in $H$ represents the titer of virus $X$ relative antiserum raised against virus $Y$. As shown above, the logarithm of the normalized titer (Normalized titer = $H^{XY}/H^{YY}$) depends additively on noise and other non-antigenic contributions. The log-transformed normalized titers corresponding to a particular virus form a vector defined in a space of dimensionality less than or equal to $n$. Basis vectors for this space can be determined by means of SVD:

$$H = U * S * V^T, \tag{S26}$$

where $U$ is an $m$ by $m$ column orthonormal matrix (i.e., $U * U^T = I_m$), $S$ an $m$ by $n$ diagonal matrix of $n$ singular values, $\lambda_i, i = 1, \ldots, n$, and $V$ an $n$ by $n$ column orthonormal matrix. The superscript $T$ denotes matrix transpose, while "*" denotes matrix multiplication. The columns of $U$ and $V$ correspond to the eigenvectors of $H * H^T$ and $H^T * H$, respectively, while the $i$th singular value in $S$ corresponds to the positive square root of the eigenvalue associated with the eigenvector found in the $i$th column of both $U$ and $V$.

The eigenvectors found in $V$ are orthonormal superpositions of the titers found in $H$. In other words, the eigenvectors represent orthogonal subsets of the variation (denoted "eigenfactors") found in titers. For illustration, consider two factors that are known to contribute to the titer of virus $X$ relative to antiserum raised against virus $Y$: the affinity of the antibodies for virus $X$ and the avidity of virus $X$ for red cells. Since these factors are, to a reasonable approximation, mutually independent their contributions to titers will be captured by different eigenfactors; this suggests a novel approach to decoupling the contributions of the mentioned factors and, therefore, to improving predictions of antigenic similarity.

The relative contribution of (or the fraction of the variation in titers that is explained by) the $i$th eigenfactor is given by:

$$p_i = \lambda_i^2 \Big/ \sum_{j=1}^{n} \lambda_j^2. \tag{S27}$$

The fraction of the variation explained by a given subset of eigenfactors is the sum of the fraction of the variation explained by the individual eigenfactors. To quantify antigenic differences between viruses, the matrix $H$ of titers is mapped onto the standard coordinate system in $R^n$ by a change of basis: $V^T * H^T$. Those eigenfactors that are deemed to reflect measurement noise (e.g., eigenfactors that make a very small relative contribution to titers, as indicated by $p_i$) are filtered out by setting the viral coordinates associated with those eigenfactors to zero. The antigenic difference (called eigendistance) between a given pair of viruses is defined as the Euclidean distance between the coordinates of those viruses, in the subspace spanned by the unfiltered eigenfactors. Broad-scale patterns of antigenic similarities among viruses are elucidated via computational clustering of the viral coordinates. If the number of unfiltered eigenfactors is small (i.e., less than or equal to 3), then the coordinates of viruses can be plotted in order to visualize the elucidated patterns. Moreover, specific subsets of the unfiltered eigenfactors can be endowed with biological significance, whenever possible, on the basis of known attributes of viruses that correlate with those eigenfactors.

In the following, two examples are used to illustrate how the above SVD method can enable the delineation of antigenically relevant variation found in titers. First, the method was applied to tables of titers for H3N2 viruses [38]. For each table, the eigenfactors were computed and all subsets of those eigenfactors of sizes ranging from 2 to 5 were determined. Eigendistances between pairs of viruses were subsequently computed for each eigenfactor subset (see above). The amount of variation explained by each eigenfactors subset was also determined (see above). The antigenic relevance of the variation explained by a particular eigenfactor subset was quantified as the average correlation between eigendistances computed using that subset and the corresponding AHM measure of the antigenic similarity between pairs of viruses. Figure S2a shows a plot of the average correlation versus the explained variation for all eigenfactor subsets and all analyzed tables. The results show that different subsets of eigenfactors can accurately predict AHM, irrespective of the amount of variation they explain (Figure S2a). An interesting feature of the results is that the ability to predict AHM increases as the explained variation also increases from 0% to ~40%, and then it drops sharply only to increase again as the explained variation increases above ~60%. The location of the sharp drop corresponds approximately to the average amount of variation explained by the first eigenfactor. This suggests that the variation explained by the first eigenfactor may not be antigenically relevant, in spite of the fact that it is the dominant eigenfactor. In contrast to the above results obtained using un-centered titers, when the titers from each table are mean-centered prior to their analysis, the relationship between the explained variation and the ability to predict AHM becomes linear (Figure S2b), suggesting that mean-centering may have minimized the effect of the first eigenfactor. This example illustrates the potential of the above method to provide insight into the nature of the variation found in titers.

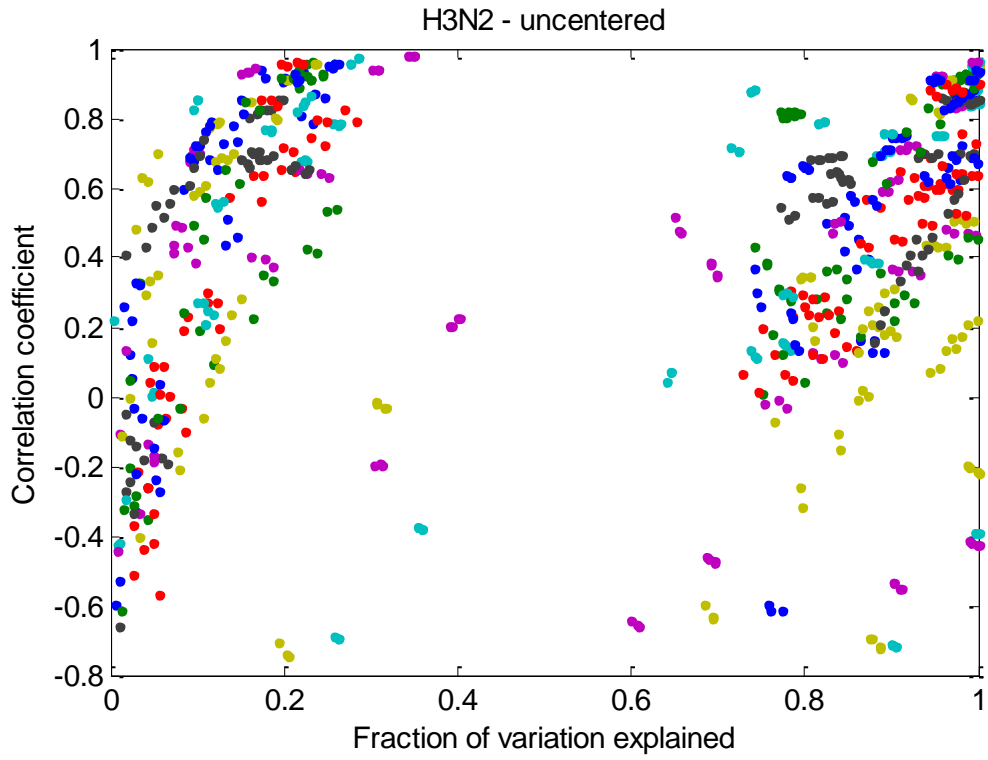

(a)

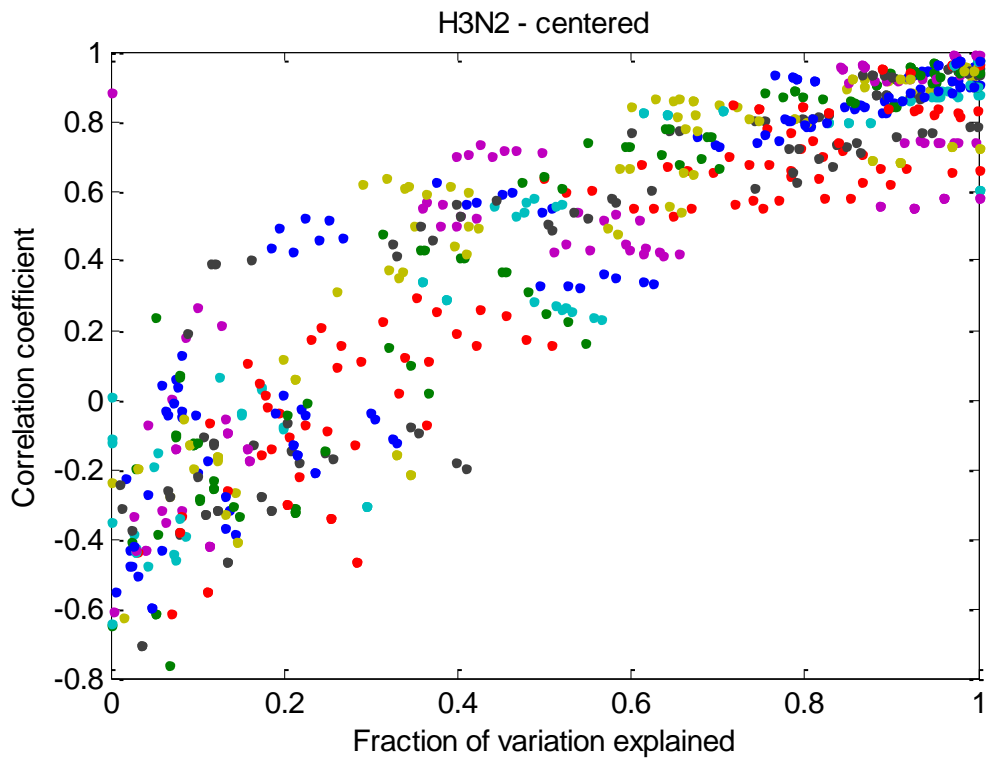

(b)

Figure S2. Predictive accuracy of subsets of eigenfactors with respect to the AHM measure of antigenic similarity. Tables of titers from [38] that contain 5 or more antisera for H3N2 viruses were analyzed. The tables were either un-centered (a) or row-centred (b) prior to their analysis. The eigenfactors for each table were computed and the viruses from each table were subsequently mapped onto subspaces spanned by each of the 26 possible subsets of the first 5 eigenfactors (excluding subsets of size 1). [The eigenfactors were ranked in decreasing order of the amount of variation explained by each]. For each table and for each subset of eigenfactors, the AHM measure of the antigenic distance between pairs of viruses as well as the eigendistances between those pairs of viruses (see above). The correlation coefficient between the negative log (base 2) of AHM and the eigendistances was plotted against the explained variation for all subsets of eigenfactors.

In addition, the SVD method was applied to a row- and column-centered version of a table of titers for 21 viruses belonging to three antigenic subtypes (H1N1, H2N2, and H3N2) of influenza viruses. The goal of this particular application was to determine the effect of mean-centering titers on the ability to recover known antigenic relationships between the viruses under consideration. A visual map of the viruses was constructed (Figure S3a), as described above. The map allowed the three natural antigenic clusters of the viruses to be accurately recovered using the *k*-means clustering algorithm [39]. In contrast, when the method was applied to versions of the above table that were either un-centered (Fig. S4b), only row-centered (Fig. S4c), or only column-centered (Fig. S3d), the resulting visual maps did not enable accurate recovery of the natural antigenic clusters of the viruses, suggesting that the proposed mean-centering procedure may be useful indeed. Note that the centered, normalized titers require at least two dimensions to be plotted. Some factors that may influence the number of required dimensions are discussed in Part 3 of this document.

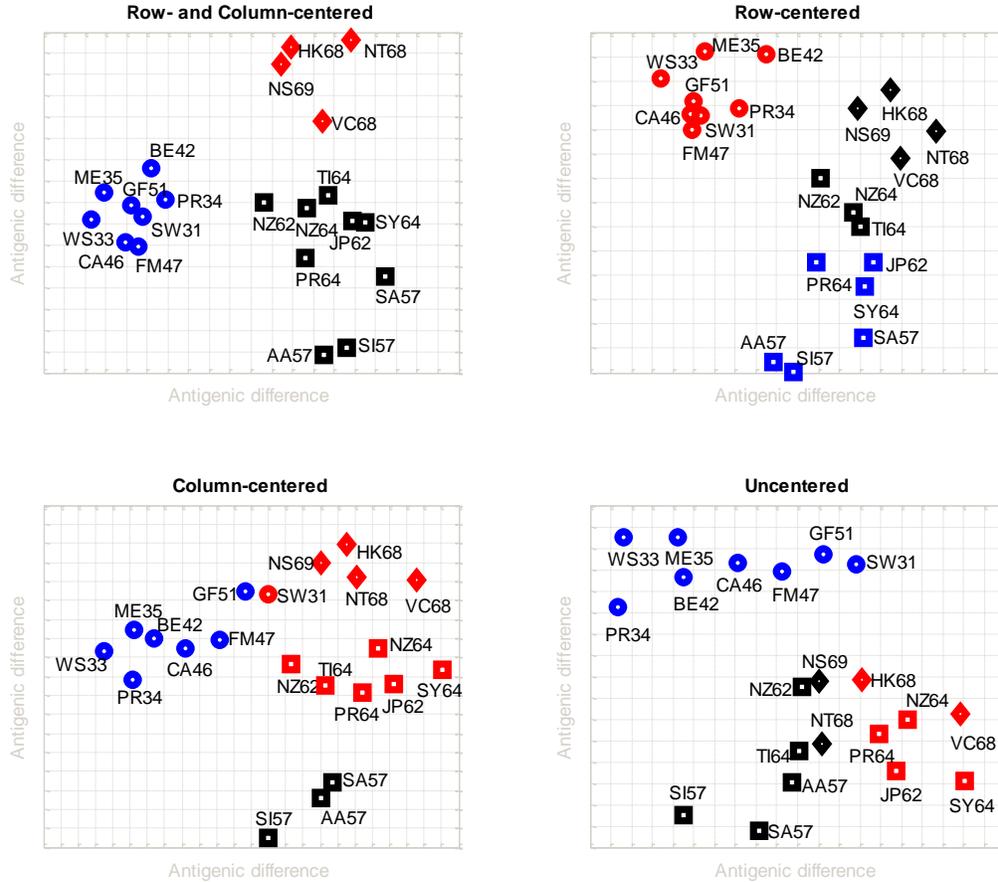

Figure S3. Antigenic maps of H1N1 (circles), H2N2 (squares), and H3N2 (diamonds) influenza viruses. The maps were constructed as described above, using a table of titers (H369001) published in [38]. The table was either un-centered, row-centered, column-centered, or row- and column-centered before it was used. Three viral antigenic clusters (colored in red, blue, and black, respectively) were obtained by applying the *k*-means clustering algorithm [39] to viral coordinates taken from each map. Virus notation: AA57 AA/23/57, BE42 Bellamy/42, CA46 Cam/46, FM47 FM/1/47, GF51 GFM/51, HK68 Hong Kong/1/68, JP62 Japan/170/62, ME35 Melbourne/35, NS69 New South Wales/5/69, NT68 Northern Territory/60/68, NZ62 New Zealand/11/62, NZ64 New Zealand/14/64, PR34 Puerto Rico/8/34, PR64 Puerto Rico/1/64, SA57 South Australia/12/57, SI57 Singapore/57, SW31 Swine/31, SY64 Sydney/2/64, TI64 A/Taiwan/64, VC68 Victoria/4/68, WS33 WSE/33.

**A general approach to SVD suitable for HI tables with missing data:** The SVD projection of a mean-centered HI table *H*, which is given by $V^T * H^T$, yields the transpose of the matrix of principal components of *H*, denoted here by the $m \times q$ matrix *W*, $q \leq n$. Indeed, the SVD projection of $H$ is recovered exactly, in the zero-noise limit, as the maximum likelihood solution of the following statistical model [40]:

$$H = W * X + \varepsilon,  \quad\quad\quad (S28)$$

where $X = V^T$ is the $q \times n$ matrix of eigenfactors, and $\varepsilon \sim N(0, \sigma^2 I)$ denotes isotropic, multidimensional Gaussian noise. It is often convenient to ensure the orthogonality of eigenfactors by requiring that $X \sim N(0, I)$. It follows that $H \sim N(0, W*W^T + \sigma^2 I)$.

The maximum likelihood estimates of $W$ and $\sigma^2$ are given by [40]:

$$W_{ML} = U_q * (\Lambda_q - \sigma^2 I)^{1/2} * R \text{ and} \quad\quad\quad (S29)$$

$$\sigma^2_{ML} = \frac{1}{m-q} \sum_{j=q+1}^{m} \lambda_j, \quad\quad\quad (S30)$$

respectively, where the columns of $U_q$ are the first $q \leq m$ eigenvectors of $H*H^T$, whose eigenvalues $\lambda_1, \ldots, \lambda_q$ are found in the diagonal matrix $\Lambda_q$. $R$ is a rotation matrix, which, for convenience, can be set to the identity matrix. In our applications to HI data, we found that $\lambda_j \sim 0$ for $j > n$, so that both the deterministic and probabilistic SVD projections are essentially equivalent in this case.

In many cases (e.g., when the viruses being compared are antigenically either very similar or very divergent) the titer is either too small or too large than can be measured reliably by the HI assay. In such cases, the titer in question is reported as a threshold value. HI tables containing such threshold (or missing) titers cannot be decomposed by using the SVD approach described in the previous section since $H*H^T$ is not well-defined in this case. Tipping and Bishop [40] proposed an expectation maximization algorithm for dealing with such missing data, but their approach requires making certain assumptions about the nature of the missing data that are untenable in the case of HI data. A better alternative is the alternating regression approach used by Liu and co-workers [41] to analyze microarray data. This approach is modified here in order to account for threshold titers.

Each entry $H_{ij}$ of a mean-centered HI table $H$ is expressed as:

$$H_{ij} = W_{i*} * X_{*j} + \varepsilon_{ij}, \quad\quad\quad (S32)$$

where $W_{i*}$ denotes the $i$th row of the matrix $W$ of principal components of $H$ and $X_{*j}$ denotes the $j$th column of the rotation matrix $X$. $W_{i*}$ and $X_{*j}$ are solved for by the following alternating regression procedure [41]:

1. Initialize $X$ (here the initial value of $X$ is drawn from $N(0,I)$).
2. Use $X$ to solve for $W$.
3. Use $W$ (obtained in step 2) to solve for $X$.
4. Repeat steps 2 and 3 until convergence of $H-W*X$.

Note that an exact solution can be obtained in steps 2 and 3 of the above algorithm. The columns of the final solution for $X$ may not be mutually orthogonal, that is, the condition $X \sim N(0,I)$ may not be satisfied. Therefore, the matrices $W$ and $X$ are transformed to ensure that this condition is satisfied:

$$H = \tilde{W} * \tilde{X} + \varepsilon, \tag{S33}$$

where $\tilde{W} = W * (G * \Sigma^{-1/2} * G^T)^{-1}$, $\tilde{X} = (G * \Sigma^{-1/2} * G^T) * X$, and $X * X^T = G * \Sigma * G^T$. It is clearly seen that $\tilde{X} \sim N(0,I)$:

$$\begin{aligned}
\tilde{X} * \tilde{X}^T &= (G * \Sigma^{-1/2} * G^T * X) * (G * \Sigma^{-1/2} * G^T * X)^T \\
&= (G * \Sigma^{-1/2} * G^T) * X * X^T * (G\Sigma^{-1/2}G^T)^T \\
&= (G * \Sigma^{-1/2} * G^T) * G * \Sigma * G^T * (G\Sigma^{-1/2}G^T)^T = I
\end{aligned} \tag{S34}$$

For a threshold titer $H_{ij}$ the expression $\exp\{\alpha \cdot sign(H_{ij}) \cdot |H_{ij} - W_{i*} * X_{*j}|\}$ is minimized, where $\alpha \geq 0$ and $sign(H_{ij})$ equals +1, if $H_{ij}$ is the lower bound of the titer, and -1, if $H_{ij}$ is the upper-bound of the titer. This expression was chosen to ensure that the error associated with $H_{ij}$ makes an appreciable contribution to the total error $\varepsilon$ [see Eqn. (S33)] only when the threshold condition represented by $H_{ij}$ is violated.

### 2.2. Probabilistic multidimensional scaling (MDS) of antigenic differences

Given a collection of HI tables, SVD is used to compute antigenic differences (or eigendistances) between the viruses found in each table (see above). These eigendistances are subsequently embedded in a common, reduced space, of dimensionality $k$, by means of probabilistic MDS [42]. In order to accomplish this, an explicit model for the noise structure of eigendistances is assumed. The distribution of the noise associated with eigendistances can be reasonably approximated by the log-normal distribution (e.g., [43]), which captures the fact that (i) titers are always non-negative, and (ii) they depend on multiplicative stochastic processes, for example, resulting from the serial dilution of sera. In other words, the likelihood of the eigendistances found in the $t$th HI table under consideration can be approximated by:

$$p(D_t|X, \sigma_t^2, k) \approx (2\pi)^{-1/2 N_t} \sigma_t^{-N_t} \prod_{i,j} \|X_{i*} - X_{j*}\|^{-1} \exp\left[\frac{-1}{2\sigma_t^2} \ln^2\left(\frac{\|X_{i*} - X_{j*}\|}{D_{ijt}}\right)\right], \quad (S35)$$

where $D_t$ denotes a matrix of eigendistances between the viruses found in the $t$th table, $\sigma_t^2$ the variance of those eigendistances, $N_t$ the number of eigendistances found in $D_t$, $\|\cdot\|$ the Euclidean norm, and $X$ a matrix of viral coordinates. The matrix $X$ has dimensions $m \times k$, where $m$ is the total number of distinct viruses found in the tables under consideration.

The maximum likelihood estimate (mle) for $\sigma_t^2$ is given by $S_t/N_t$, where

$$S_t = \sum_{i,j} \log^2\left(\frac{\|X_{i*} - X_{j*}\|}{D_{ijt}}\right). \quad (S36)$$

Substituting the mle for $\sigma_t^2$ into (S35) gives the joint log-likelihood function (excluding terms that do not depend on parameters) of eigendistances from all the tables under consideration:

$$\log L_k = -\sum_t \frac{N_t}{2}\left(\log S_t + 1 - \log N_t\right), \quad (S37)$$

where the subscript emphasizes the dependence on $k$. It follows from (4) that $L_k$ attains its maximum value, $\hat{L}_k$, when $\sum_t N_t \log S_t$ is minimized with respect to $X$; this also yields the mle for $X$, which is subsequently mean-centered and transformed to ensure that its columns are mutually orthogonal[9]. $\hat{L}_k$ is used to estimate the optimal value of $k$, defined as the value of $k$ that minimizes the Bayesian information criterion [44]: $-\hat{L}_k + \frac{1}{2}(k \cdot m)\log(\sum_t N_t)$. Note that $\sum_t N_t \log S_t$ is minimized using the method of simulated annealing [45], since this method can circumvent the local minima that are characteristic of the solution of space of this type of nonlinear optimization problem.

---

[9] If we transform X by setting $X = X * U$, where $U * S * U^T$ is the SVD of $X^T * X$, then the columns of $X$ will become mutually orthogonal: $X^T * X = S$.

**Confidence regions for the MDS coordinates of viruses:** After obtaining the mle, $\hat{X}_0$, for the $m \times k$ matrix of viral coordinates from the MDS of eigendistances, 95% confidence regions for these coordinates are computed as follows: 100 bootstrap copies of the eigendistances are created, and the mle, $\hat{X}_i$, for viral coordinates consistent with the $i$th bootstrap copy, $1 \leq i \leq 100$, is computed. Each $\hat{X}_i$ is then mean-centered and transformed (via a Procrustrean fit) so that its Euclidean distance to $\hat{X}_0$ is minimized. Specifically, the following transformation is applied to each $\hat{X}_i$ [46]:

$$\hat{X}_i^* = \text{trace}(\hat{X}_i * V \cdot U^T * \hat{X}_0^T)/\text{trace}(\hat{X}_i * \hat{X}_i^T) * \hat{X}_i * V * U^T, \tag{S38}$$

where $U * S * V^T$ is the SVD of $\hat{X}_0^T * \hat{X}_i$. The centroid and the covariance matrix of the "bootstrap" coordinates of each virus (extracted from $\hat{X}_i^*$, $i = 1 \cdots 100$) are used to estimate 95% confidence regions for viral coordinates.

## Part 3. Effective dimensionality and recovery of titers
### 3.1. Factors that may affect the effective dimensionality of titers

Let $H = U * S * V^T$ be the SVD of an $m \times n$ table $H$ of log-transformed titers. Recall that the columns of $U$ ($V$) are called eigenvectors, and the diagonal entries of $S$ are the $n$ singular values of $H$, $\lambda_i, i = 1,\ldots,n$, sorted in decreasing order of magnitude. The rank (or dimensionality) of $H$ is the number of its non-zero singular values. When all the $n$ singular values are non-zero, $H$ is said to be of full rank. If the rank of $H$ is $r \leq n$ and its entries are not contaminated by noise, then only the first $r$ $\lambda$'s will be non-zero. However, if the entries are contaminated by noise, then some of the other $n$-$r$ $\lambda$'s can also be non-zero. In this case, it is necessary to distinguish the $\lambda$'s that are non-zero due to the "natural" variation of the data and those that are non-zero due to noise. A number of statistical methods have been developed to address this problem. Among the most popular of these are methods that make certain assumptions about the distribution of $(\lambda_i)^2, i = 1, \ldots, n$, the eigenvalues of $H^T H$ [47]; each eigenvalue is said to represent the amount of variation explained by its corresponding eigenvector [47]. Common assumptions made by those methods include: (i) the noise contaminating the entries of $H$ is normally distributed, as are the eigenvalues, and (ii) the noise-associated eigenvalues are significantly smaller than the other eigenvalues, since the latter eigenvalues reflect contributions from both noise and the natural variation of the data. Based on these assumptions, the "true" non-zero eigenvalues can be identified by determining whether each eigenvalue is significantly greater than the mean of the eigenvalues smaller than it. This can be done by finding the maximum value of *r* for which:

$$\frac{(\lambda_r)^2}{\sum_{i=r+1}^{n}(\lambda_i)^2}(n-r) > f_{1,n-r}(1-\alpha), \tag{S39}$$

where $f_{1,n-r}$ denotes the inverse cumulative function of the *F*-distribution with 1 and *n-r* degrees of freedom, and $\alpha$ is the level of statistical significance of the test. The above method for determining the effective rank, commonly called Malinowski's *F*-test, was developed by using Fisher's variance-ratio test [47]. An equivalent method was also developed by Carey et al. [48] by using a different approach. Note that (S39) is equivalent to Eqn. (3) of the main text.

Note that a number of different factors can cause variation in estimates of the effective dimensionality of titers. Some of those factors are discussed below.

First, possible biological bases of the dimensions of titers are considered. The dimensions of titers may represent independent attributes of viruses and sera that determine the outcomes of the biophysical interactions that take place during the HI assay (see Part 1 for details on these interactions). In particular, the variation in titers arises from, among other factors, variation in viral attributes such as the affinity for red cell and susceptibility to antibody-mediated neutralization, and from attributes of sera such as the concentration of antibodies and the affinity of those antibodies for virus. Some of these attributes may vary independently of others and they may therefore form distinct dimensions of titers. Regardless of the specific nature of the attributes that constitute the individual dimensions of titers, the relative contributions of those attributes to the variation in titers can have an important influence on the effective dimensionality. In particular, if the variation due to some attributes is substantially greater than the variation due to others, then the effective dimensionality can be lower than the actual dimensionality depending on the number of those "dominant" attributes. Moreover, if the contribution of the different attributes to variation in titers depends on the particular viruses and sera used in a given experiment, then this may cause the effective dimensionality of titers from independent experiments to also vary.

Another factor that can affect the effective dimensionality of a given table of log-transformed titers $H$ is the way that the titers are processed. Consider, for example, the titer-normalization procedure of [1]. Mathematically, that procedure can be expressed as:

$$\tilde{H} = -H + \vec{1}\vec{v}, \tag{S40}$$

where $\tilde{H}$ is an $m \times n$ matrix of normalized titers, $\vec{1}$ an $m \times 1$ vector of ones, and $\vec{v}$ a $1 \times n$ vector whose entries are the maximum values found in each of the $n$ rows of $H$. If $H$ is of rank $r$, then it can be written as the product of two matrices each of rank $r$: $H = PQ$, where $P$ has dimensions $m \times r$ and $Q$ has dimensions $r \times n$ [49]. Therefore, $\tilde{H}$ can be rewritten as:

$$\tilde{H} = \begin{pmatrix} -P & \vec{1} \end{pmatrix} \begin{pmatrix} Q \\ \vec{v} \end{pmatrix}. \tag{S41}$$

If $\vec{1}$ does not belong to the column space of $P$, then the matrix $\begin{pmatrix} -P & \vec{1} \end{pmatrix}$ will be of rank $r+1$ (e.g., see [48]). Similarly, if $\vec{v}$ does not belong to the row space of $Q$, then $\begin{pmatrix} Q \\ \vec{v} \end{pmatrix}$ will be of rank $r+1$. Therefore, titer normalization will increase the rank of $H$ by 1 if both of the above conditions hold, since [49]:

$$r(\tilde{H}) = \min\left\{ r\left[\begin{pmatrix} -P & \vec{1} \end{pmatrix}\right], r\left[\begin{pmatrix} Q \\ \vec{v} \end{pmatrix}\right] \right\}. \tag{S42}$$

In addition, the rank of a table of titers can increase when titers from different experiments are combined. For example, consider the case when titers for $m$ viruses are measured in two experiments, the first involving $n_1$ sera and the second involving $n_2$ sera. Denote by $H_1$ ($H_2$) the table of titers obtained from the first (second) experiment. The dimensions of $H_1$ and $H_2$ will be $m \times n_1$ and $m \times n_2$, respectively. We are interested in determining the rank of the $m \times (n_1 + n_2)$ composite table $C = (H_1, H_2)$, consisting of titers measured in the two experiments. It is well known that the rank of $C$ is given by [49]:

$$r(C) = r(H_1) + r[H_1, (I - H_1^- H_1)H_2] \geq r(H_1) + \min\{r(H_1), r[(I - H_1^- H_1)H_2]\}, \qquad (S43)$$

where $H_1^-$ is the generalized inverse of $H_1$ and $I$ is the identity matrix. $H_1^-$ can be obtained from the SVD of $H_1$, $H_1 = USV^T$, as follows: $H_1^- = V\hat{S}^T U^T$, where $\hat{S}$ is obtained from $S$ by replacing each non-zero diagonal entry by its reciprocal. It follows from (S43) that the rank of $C$ will be greater than that of $H_1$ if $(I - H_1^- H_1)H_2$ does not have a rank of 0. Moreover, if the rank of $H_1$ is greater than 1, then the rank of $C$ will also be greater than 1 irrespective of the rank of $H_2$. This is also true for an arbitrary number $s$ of tables of titers measured for the same $m$ viruses; to see this, simply partition the tables into two sub-tables, $H_1$ and $(H_2, \ldots, H_s)$. The above statement is also true if $H_2$ contains titers for an additional $m_2$ viruses that are different from the $m$ viruses found in $H_1$. In this case, $C$ can be written as:

$$C = \begin{pmatrix} H_1 & H_2 \\ O & \end{pmatrix}, \qquad (S44)$$

where $O$ is a matrix of zeros. $O$, $C$, and $H_2$ have dimensions $m_2 \times n_1$, $(m+m_2) \times (n_1+n_2)$, and $(m+m_2) \times n_1$, respectively. Following the same approach used above, the rank of $C$ can be expressed as:

$$r(C) = r\begin{pmatrix} H_1 \\ O \end{pmatrix} + r\left[\begin{pmatrix} H_1 \\ O \end{pmatrix}, \left(I - \begin{pmatrix} H_1 \\ O \end{pmatrix}^- \begin{pmatrix} H_1 \\ O \end{pmatrix}\right) H_2\right] = r(H_1) + r[H_1, (I - H_1^- H_1)H_2], \qquad (S45)$$

where we have used the fact that [49]:

$$\begin{pmatrix} H_1 \\ O \end{pmatrix}^- = \begin{pmatrix} O & H_1^- \end{pmatrix}. \qquad (S47)$$

Eqn. (S45) is identical to (S43); the conclusions reached above also apply here.

### 3.2. Nuclear-norm minimization for the recovery of unmeasured titers

Given an $m \times n$ table $H$ consisting of log-transformed titers, titers missing from the table ("unmeasured" titers) are recovered by finding an $m \times n$ matrix $X$ that minimizes:

$$\mu \|X\|_* + \eta \sum_{H^{ij} \in \Omega} \left(H^{ij} - X^{ij}\right)^2 , \tag{S48}$$

where $\mu$ and $\eta$ are regularizing constants and $\|X\|_*$ denotes the nuclear norm of $X$, the sum of the singular values of $X$. This optimization problem was solved using the fixed point continuation (FPC) algorithm of Ma et al. [50], with default parameters and two minor modifications. Firstly, the algorithm was modified to use exact SVD rather than the approximate SVD implemented in the original algorithm. This is because the dimensions of the tables of empirical titers analyzed in this study were small enough that exact SVD of those tables was computationally very efficient. Secondly, the algorithm was modified to allow the minimization of only the $r$ largest singular values of $X$ (rather than $\|X\|_*$) in cases when $r$ is known. Note that in the FPC algorithm, η is set to a theoretically motivated value of ½ [50], whereas an initial value of μ is chosen to accord $\|X\|_*$ a similar weight as the second sum of (S48). The chosen value of μ is then iteratively decreased. The Matlab code used in this study is available upon request. It will eventually be incorporated into software for analyzing serological data that will be made freely available to the public.

**Table S1. List of tables of empirical titers analyzed in this study**

```
H308001,  H307002,  H305002,  H304001,  H302001,  H302002,  H301001,  H301002,
H301003,  H300002,  H30003,   H300004,  H399001,  H399002,  H387001,  H375001,
H369001,  H108001,  H107001,  H104001,  B001001,  B001002,  B001003,  B000001
```

The listed tables are a subset of the tables published in Supplement A of [38], which only contain known titers -- titers that occur in the interval (10 to 10240). Because it was of interest to compute the first five singular values of each table, only tables containing at least five columns are listed.